\documentclass{article}
\usepackage{spconf,amsmath,graphicx}
\usepackage{multirow}
\usepackage{amssymb}
\usepackage{booktabs}
\usepackage{marvosym}

\title{ROBUST WAKE WORD SPOTTING WITH FRAME-LEVEL CROSS-MODAL ATTENTION BASED AUDIO-VISUAL CONFORMER}
\name{Haoxu Wang$^{1}$, Ming Cheng$^{1}$, Qiang Fu$^{2}$, Ming Li$^{1\dagger}$\thanks{$\dagger$ Corresponding Author, E-mail: 	ming.li369@dukekunshan.edu.cn} }
\address{
$^{1}$Suzhou Municipal Key Laboratory of Multimodal Intelligent Systems,\\ Duke Kunshan University, Kunshan, China \\
$^{2}$Research Center for Intelligent Robotics, Research Institute of Interdisciplinary Innovation, \\ Zhejiang Laboratory, Hangzhou, China \\}

\begin{document}
\ninept
\maketitle
\begin{abstract}
In recent years, neural network-based Wake Word Spotting achieves good performance on clean audio samples but struggles in noisy environments. Audio-Visual Wake Word Spotting (AVWWS) receives lots of attention because visual lip movement information is not affected by complex acoustic scenes. Previous works usually use simple addition or concatenation for multi-modal fusion. The inter-modal correlation remains relatively under-explored. In this paper, we propose a novel module called Frame-Level Cross-Modal Attention (FLCMA) to improve the performance of AVWWS systems. This module can help model multi-modal information at the frame-level through synchronous lip movements and speech signals. We train the end-to-end FLCMA based Audio-Visual Conformer and further improve the performance by fine-tuning pre-trained uni-modal models for the AVWWS task. The proposed system achieves a new state-of-the-art result (4.57\% WWS score) on the far-field MISP dataset.


\end{abstract}
\section{Introduction}
\label{sec:intro}
\vspace{-3pt}

Wake word spotting (WWS), also called Keyword Spotting (KWS), is vital in speech signal processing. It aims to detect specific keywords or phrases in audio streams. It is crucial for voice-activated devices like smart speakers, mobile phones, and virtual assistants.
Recently, deep neural networks have been used \cite{Sun2017CompressedTD} for good performance in near-field clean speech environments, such as Convolution Neural Network (CNN) \cite{Sainath2015ConvolutionalNN} and Transformer \cite{kwsTransformer}. However, the performance of these systems may significantly degrade in far-field settings due to complex environments, e.g. speech overlap, background noise and etc. Some methods aim to enhance the noise robustness of WWS systems. For instance,  researchers propose domain aware
training systems \cite{wu20j_interspeech}, sample generation \cite{wang22c_odyssey}, and multi-look minimum variance distortion less response (MVDR) beamformers \cite{9747084} to improve performance in far-field and complex scenarios.





\begin{figure}[t]
    \centering
    \includegraphics[width=0.33\textwidth]{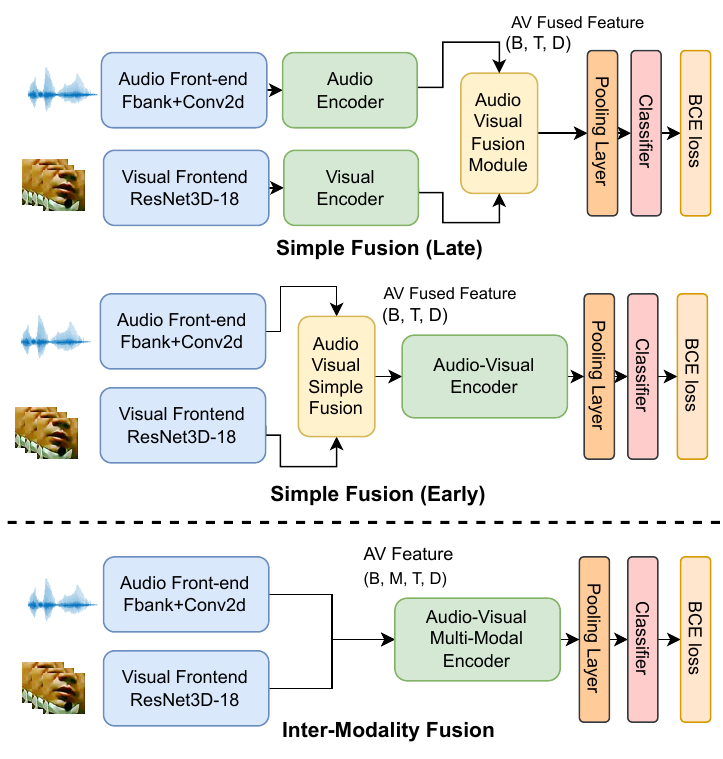}
    \vspace{-15pt}
    \caption{Diagram of simple fusion method (Late: top, Early: middle) and our proposed inter-modality fusion method (bottom). $B$, $T$ and $D$ denote the batch size, the number of frames and the dimension of the embeddings. $M$ represents the number of modalities.}
    \label{framework}
    \vspace{-18pt}
\end{figure}


Because visual lip movement information is not affected by acoustic noise and can serve as complementary information to the audio stream, the multi-modal audio-visual systems have become more and more popular  in several fields, including automatic speech recognition (ASR) \cite{avconformer, aveffconformer,avhubert2}, speech separation \cite{avsp1}, speaker verification (SV) \cite{avspeaker1}, and 
etc. Audio-Visual systems show improved performance in high-noise environments compared to audio-only systems.
\cite{seewakewords} develops a new audio-visual KWS system based on CNN. Along with the first Multimodal Information Based Speech Processing Challenge (MISP Challenge 2021 \cite{mispchallenge}) and its data release \cite{zhou22g_interspeech}, many new research works are reported targeting the Audio Visual Wake Word Spotting (AVWWS). \cite{cheng2022dku} proposes a CNN-3D-based model, and \cite{beike} proposes a transformer-based model. \cite{vekws} and \cite{dkuv2} improve their systems to enhance performance further based on \cite{cheng2022dku}. However, previous works \cite{cheng2022dku,vekws,dkuv2} usually train two robust single-modality models and then fuse them. They do not use the end-to-end (E2E) strategy to optimize the network of two modalities simultaneously. \cite{beike} trains an audio-visual E2E Transformer model but still uses single-modality models to vote the final results. Moreover, \cite{avconformer,aveffconformer,hong22_interspeech} in the audio-visual speech recognition (AVSR) domain fuse the multi-modal information using simple addition or concatenation strategy. 

Inspired by \cite{wang2022cross, horiguchi2022multi, yu2023mfcca}, which enhance multi-channel speaker diarization and ASR using Channel-Level Cross-Channel Attention (CLCCA) for frame-level correlation modeling of multi-channel speech signals, we propose the Frame-Level Cross-Modal Attention (FLCMA) module to improve the performance of the AVWWS system. Instead of simple addition or concatenation fusion, FLCMA models multi-modal semantic information at the frame level. We adopt an E2E training strategy, training an AVWWS Conformer for two modalities simultaneously to enhance system performance. Additionally, we utilize a pretrain strategy, pre-training robust uni-modal models, transferring their parameters to the multi-modal model, and performing fine-tuning. This approach achieves a new state-of-the-art (SOTA) result (4.57\% WWS score) on the MISP dataset, showcasing the effectiveness of our E2E AVWWS system.

\begin{figure}[t]
    \centering
    \includegraphics[width=0.40\textwidth]{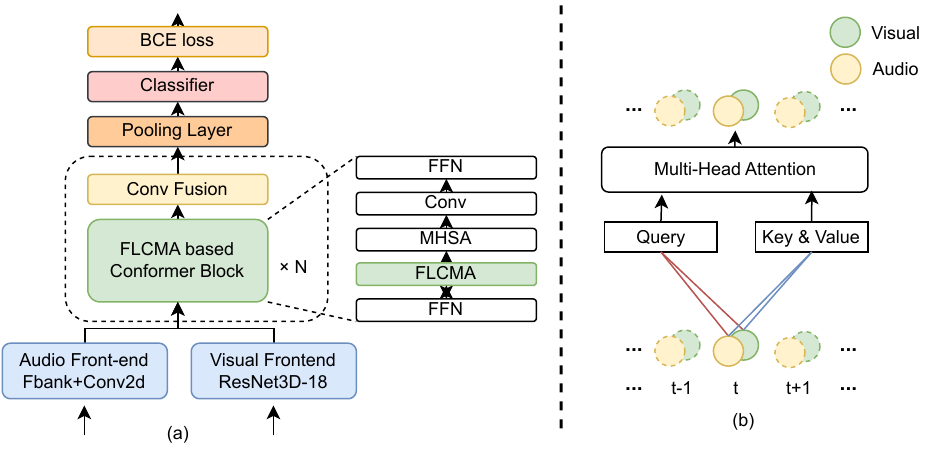}
    \vspace{-13pt}
    \caption{(a) Framework of our FLCMA based audio-visual conformer wake word spotting system. (b) Diagram of FLCMA module. Each circle means audio or visual feature embedding at each time step.}
    \label{methods}
    \vspace{-15pt}
\end{figure}

\vspace{-7pt}
\section{Methods}
\label{sec:sec2}
\vspace{-6pt}

In this section, we introduce the Frame-Level Cross-Modal Attention module. 
As shown in Fig.\ref{framework}, assumed that we extract audio feature $X_{spec} \in \mathbf{R}^{T \times D}$ and visual feature $X_{lip} \in \mathbf{R}^{T \times D}$ of $T$ frames, previous works usually get the audio-visual early fused feature by concatenating along feature dimension or adding, in which $X_{fused} = [X_{spec} || X_{lip}]W_{fc}$ or $X_{fused} = X_{spec} + X_{lip}$, where $||$ means concatenation and $W_{fc} \in \mathbf{R}^{2D \times D}$ means the fully-connected projection layer. These usually are called Simple Fusion (Early). Moreover, some works late fuse the multi-modal features after the single-modality encoder, usually called Simple Fusion (Late). These approaches only involve a simple fusion of multi-modal features and thus do not effectively model the correlations among multi-modal information. Thus, we propose the FLCMA module. We expand a new modal dimension for each single modality feature and define the multi-modal feature as $X_{av} = [X_{spec}, X_{lip}] \in \mathbf{R}^{M \times T \times D}, M = 2$ by concatenating along the new modal dimension. 

\vspace{-7pt}
\subsection{Frame-Level Cross-Modal Attention}
\vspace{-3pt}

FLCMA focuses on modeling the inter-modality information on each time frame. As the Figure \ref{methods}(b) shows, the attention module is defined as:

\vspace{-15pt}

\begin{equation}
\begin{array}{cc}
    Q_{i}^{cm} = X_{av}W_{i}^{cm,q} + (b_{i}^{cm,q})^{T} \in \mathbf{R}^{T \times M \times \frac{D}{h}}, \\
    K_{i}^{cm} = X_{av}W_{i}^{cm,k} + (b_{i}^{cm,k})^{T} \in \mathbf{R}^{T \times M \times \frac{D}{h}}, \\
    V_{i}^{cm} = X_{av}W_{i}^{cm,v} + (b_{i}^{cm,v})^{T} \in \mathbf{R}^{T \times M \times \frac{D}{h}}, \\
    O_{i}^{cm} = softmax(\frac{Q_{i}^{cm}(K_{i}^{cm})^{T}}{\sqrt{\frac{D}{h}}})V_{i}^{cm} \in \mathbf{R}^{T \times M \times \frac{D}{h}} \\
    O^{cm} = [O_{0}^{cm}, O_{1}^{cm}, ..., O_{h}^{cm}] \in \mathbf{R}^{T \times M \times D} \\
\end{array}
\end{equation}

where $Q_{i}^{cm}, K_{i}^{cm}, V_{i}^{cm}$ are the i-th head of Query, Key, Value of this attention module, $W_{i}^{cm,*}, b_{i}^{cm,*}$ are learnable parameters, and $h$ is the number of attention heads. This FLCMA module can help capture inter-modality correlations at the frame level through the high synchronous lip movements and speech signal. By leveraging multi-modal information, this module provides complimentary lip movement for noisy acoustic information and reduces phoneme confusion for lip movement according to the speech signal.

\vspace{-7pt}
\subsection{Pretrain Strategy}
\vspace{-3pt}


The Pretrain strategy, widely employed in speech recognition \cite{hsu2021hubert} and WWS tasks \cite{cheng2022dku,dkuv2}, has demonstrated its effectiveness. It involves pretraining an unsupervised model on an unlabeled database and fine-tuning it on a labeled database. Some approaches are to pretrain on one labeled database and fine-tune on another labeled database, known as transfer learning. In this work, we adopt the pretrain strategy by training single-modal models with a similar architecture to the multi-modal model. We transfer the parameters from the corresponding single-modal models to the multi-modal model, allowing it to leverage the knowledge acquired from the single-modal models. This approach provides a strong starting point for the multi-modal model and enhances its performance.

\vspace{-7pt}
\subsection{E2E Model Architecture}
\vspace{-3pt}

Here, we present our AVWWS system, the FLCMA based Audio-Visual Transformer or Conformer. As shown in Figure \ref{methods}(a), our system consists of 6 modules: audio and visual frontend, Transformer or Conformer encoder with FLCMA module, convolution fusion module, attentive pooling layer and the final fully-connected classifier.

\vspace{-8pt}
\subsubsection{Front-end}
\vspace{-3pt}

For the visual stream, a ResNet-18 with 3D convolution layers is employed to transform the input video frames into temporal features. The visual inputs have a shape of (T, H, W, C), which is then squeezed along the spatial dimension using global average pooling to obtain a shape of (T, D). A linear layer is then used to project these temporal features to the dimension of the conformer encoders.


For the audio stream, a subsampling module comprising two 2D convolution layers is used to convert the extracted acoustic features into temporal features. A linear layer is then employed to project these features to the dimension of the encoders. As the video's sampling rate is generally lower than that of audio, the audio features are downsampled in the time dimension to $\frac{T'}{rate}$ after standard feature extraction, aligning the time dimension of audio and video ($\frac{T'}{rate} = T$). Here we get the audio and visual features $X_{spec}$, $X_{lip}$.

\vspace{-8pt}
\subsubsection{Encoder}
\vspace{-3pt}
We use the modified transformer or conformer block as our encoder layer. FLCMA based Transformer block includes a FLCMA module, a standard multi-head self-attention (MHSA) module, and a feed-forward (FFN) module. FLCMA based Conformer block includes a FLCMA module, a MHSA module, a convolution (CONV) module, and a pair of FFN modules in the Macaron-Net style. Standard conformer block combines CONV and MHSA to capture local and global correction in the single modality. Together with the FLCMA module, this modified conformer block can further leverage the inter-modality information at the frame level.




\begin{figure}[ht]
    \centering
    \vspace{-7pt}
    \includegraphics[width=0.37\textwidth]{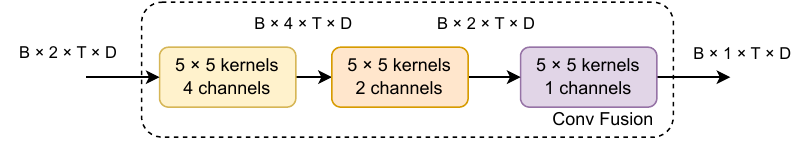}
    \vspace{-10pt}
    \caption{Framework of convolution fusion module.}
    \label{convfusion}
    \vspace{-20pt}
\end{figure}

\vspace{-2pt}
\subsubsection{Convolution fusion}
\vspace{-3pt}

Inspired by \cite{yu2023mfcca}, we use a convolution fusion module to fuse the audio-visual feature after the conformer encoder blocks instead of averaging or concatenation, which is usually used in previous works \cite{avconformer, beike}. As shown in Fig \ref{convfusion}, We use a multi-layer convolution module to help reduce the corruption caused by the direct fusion of multi-modal features, which consists of a series 2D convolution module with the channel of 4, 2, 1.

\vspace{-8pt}
\subsubsection{Attentive Pooling and Classifier}
\vspace{-3pt}


Additionally, we incorporate an attentive pooling layer, commonly utilized in SV, to capture the significance of each frame and extract a more robust classification vector. This vector is then passed through a series of fully-connected linear layers with a sigmoid function to output the probability of the wake word.

\vspace{-7pt}
\section{Experimental Setup}

\vspace{-7pt}
\subsection{Dataset and Evaluation Metrics}
\vspace{-3pt}


We evaluate our proposed system on the AVWWS dataset from the 1st MISP challenge 2021 \cite{mispchallenge}. This dataset is utilized to detect the wake word 'Xiao T, Xiao T' spoken in far-field home scenarios.
The released database contains about 125 hours and has two subsets: training set (47k+ negative samples and 5K+ positive samples) and development (Dev) set (2k+ negative samples and 600+ positive samples). Audio samples include single-channel near-field audio, 2-channel middle-field audio, and 6-channel far-field audio; video samples include single-person high-definition middle-field and multi-person far-field video. Moreover, an evaluation set (8K +) without annotations is provided to competition participants, which is only in the far-field. We obtain the annotations from the MISP committee to ensure a fair comparison of our results with those of other teams.


To evaluate our system's performance, we follow the guidelines provided by the competition committee. We utilize the False Reject Rate (FRR), False Alarm Rate (FAR), and the WWS Score. Let $N_{wake}$ represent the number of samples containing the wake word, and $N_{non\_wake}$ represent the number of samples without the wake word. The FRR and FAR are defined as: 
\vspace{-7pt}

\begin{equation}
FRR = \frac{N_{FR}}{N_{wake}}, \quad FAR = \frac{N_{FA}}{N_{non\_wake}}
\end{equation}
\vspace{-7pt}

\noindent where $N_{FR}$ denotes the number of samples containing the wake word while not recognized by the system. $N_{FA}$ denotes the number of samples containing no wake words while predicted to be positive. Hence, the final score of Wake Word Spotting (WWS) is defined as: 
\vspace{-7pt}
\begin{equation}
Score^{WWS} = FRR + FAR
\end{equation}

\vspace{-12pt}
\subsection{Experimental Setup}
\vspace{-3pt}

\textbf{Preprocess}: For the audio stream, we extract log-mel filterbank features (FBank) with a dimension of 80 from the waveform using a window length of 25ms and a shift of 10ms. Each audio sample is sampled to contain 256 frames, which means that $T^{'}$ is set to 256. Furthermore, after the audio front-end module, the time dimension of the audio features is downsampled to a quarter of its original size and is aligned to the lip movements.

For the visual stream, we refer to \cite{cheng2022dku}, focus solely on the RGB lip region images of the video. We use a face detector (RetinaFace \cite{deng2019retinaface}) and a face recognizer (ArcFace \cite{deng2019arcface}) to get the reference face in the far-field video and the face landmarks. We crop the lip regions of the target speaker from the facial images based on the detected facial landmarks according to \cite{cheng2022dku,yang2019lrw}. The lip-region videos are extracted and resized to a resolution of $112\times112$ with 3 RGB channels. The videos are sampled to contain 64 frames, resulting in a shape of $(64,112,112,3)$ and meaning $T$ is set to 64. Additionally, each pixel value in the video is normalized to the range of $[0,1]$.


\noindent \textbf{Data Augmentation}:
For the audio stream, we use various techniques inspired by \cite{mispchallenge, cheng2022dku}, including negative sub-segmentation, speed perturbation, slight trimming, and SpecAugment \cite{specaugment}. We also perform several additional steps on the original near-field audio to simulate middle and far-field audio. The pyroomacoustic tool is used to generate room impulse responses. We also incorporate noises provided by official sources, randomly adjusting the signal-to-noise ratio (SNR) within the range of -15 to 15 dB. The MVDR method is also used to add beamforming-enhanced audio into our training data. 

For the visual stream, we also use the same video-based data augmentation methods referred to the \cite{cheng2022dku}, including speed perturbation, frame-wise rotation, horizontal flip, frame-level cropping, color jitters and gray scaling. Additionally, we incorporate random histogram equalization to augment the data further. 


\noindent \textbf{Model Training}: For the FLCMA module based transformer or conformer structure, we use 6 self-attention blocks, each with 4 heads, 256-dimensional hidden size, and 1,024 dimensions for the feed-forward layer, which means that $D = 256, h = 4, N = 6$. The batch size is set to 48. The learning rate is set to 0.001 and warmed up at the first 10,000 steps by the Adam optimizer. And we adopt the weighted BinaryCrossEntropy (BCE) Loss (negative:positive=1:5) to tackle the imbalance between positive and negative samples. For the Pretrain strategy, we first train two uni-modal models, which consist of a frontend module, uni-modal encoder, pool layer and classifier using the above setup, then initialize our FLCMA based multi-modal model with the parameters of the uni-modal models, finally fine-tune the multi-modal model with the learning rate of 0.0001.



\begin{table}[t]
\small

\setlength{\tabcolsep}{0.8mm}{
  \centering
  \footnotesize
  \caption{\label{ablationtable} {\it Ablation study results of our proposed FLCMA module and Pretrain strategy. For the sake of simplicity of presentation, all AUCs have 99.0 as the 0.0 baseline, i.e. 0.636 means 99.636. L means Late, and E means Early.
  \vspace{+0.2em}
}}
  \begin{tabular}{lcccccccc}
  \toprule
  \multirow{2}*{\textbf{Methods}} & \multicolumn{4}{c}{\textbf{Dev[\%]}} & \multicolumn{4}{c}{\textbf{Eval[\%]}}  \\
  \cmidrule(lr){2-5} \cmidrule(lr){6-9}   & AUC & FRR & FAR & WWS & AUC & FRR & FAR & WWS \\
  
  \midrule
  AV-Transformer(L) & 0.585 & 3.69 & 2.02 & 5.71 & 0.421 & 4.23 & 2.44 & 6.67 \\
  AV-Transformer(E) & 0.232 & 2.72 & 4.76 & 7.48 & 0.083 & 1.66 & 7.29 & 8.95 \\
  \ \ + FLCMA & 0.453 & 1.76 & 4.18 & 5.94 & 0.518 & 1.04 & 6.02 & 7.06 \\
  \ \ \ \ + Pretrain & 0.645 & 2.24 & 2.21 & 4.45 & 0.576 & 2.14 & 3.33 & 5.47 \\
  AV-Conformer(L) & 0.510 & 5.45 & 1.68 & 7.13 & 0.416 & 4.54 & 2.16 & 6.70 \\
  AV-Conformer(E) & 0.456 & 4.97 & 1.92 & 6.89 & 0.231 & 4.78 & 2.99 & 7.77 \\
  \ \ + FLCMA  & 0.617 & 2.24 & 2.31 & 4.55 & 0.541 & 1.59 & 3.91 & 5.50 \\
  \ \ \ \ + Pretrain & \textbf{0.699} & 2.08 & 1.78 & \textbf{3.86} & \textbf{0.636} & 2.02 & 2.55 & 
\textbf{4.57} \\

  \bottomrule

\end{tabular}
}
\vspace{-15pt}
\end{table}

\begin{table}[t]
\small

\setlength{\tabcolsep}{1.3mm}{
  \centering
  \footnotesize
  \caption{\label{audiotable} {\it Comparisons of recent released uni-modal systems.
  \vspace{+0.2em}
}}
  \begin{tabular}{lcccccccc}
  \toprule
  \multirow{2}*{\textbf{Methods}} & \multicolumn{3}{c}{\textbf{Dev[\%]}} & \multicolumn{3}{c}{\textbf{Eval[\%]}}  \\
  \cmidrule(lr){2-4} \cmidrule(lr){5-7}   & FRR & FAR & WWS & FRR & FAR & WWS \\
  
  \midrule
    V-LSTM \cite{zhou22g_interspeech} & 38.4 & 8.7 & 47.1 & 31.7 &  26.7 & 58.4 \\
    V-ResNet3D \cite{cheng2022dku} & 8.65 & 8.41 & 17.06 & - & - &  21.7 \\
    V-ResNet3D \cite{vekws} & 6.73 & 6.68 & \textbf{13.41} & 18.39 & 7.67 & 26.01 \\
    V-SimAM \cite{dkuv2} & 9.13 & 6.25 & 15.39 & 8.03 & 11.1 & \textbf{19.13} \\
    V-Transformer (Ours) & 12.18 & 5.29 & 17.47 & 15.63 & 7.45 & 23.08 \\
    V-Conformer (Ours) & 10.41 & 5.58 & 15.99 & 13.36 & 7.94 & 21.30 \\
    \midrule
    A-LSTM \cite{zhou22g_interspeech} & 10.4 & 6.0 & 16.4 & 14.7 & 11.5 & 26.2 \\
    A-ResNet3D \cite{cheng2022dku} & 6.41 & 6.01 & 12.42 & - & -& 12.2 \\
    A-Conformer \cite{beike} & - & - & 10.5 & - & - & 11.6 \\
    A-ResNet3D \cite{vekws} & 6.73 & 3.12 & 9.85 & 5.82 & 3.95 & \textbf{9.78} \\
    A-SimAM \cite{dkuv2} & 5.93 & 3.61 & \textbf{9.54} & 6.38 & 4.65 & 11.03 \\
    A-Transformer (Ours) & 4.81 & 10.05 & 14.86 & 3.55 & 9.04 & 12.59 \\
  A-Conformer (Ours) & 5.28 & 6.88 & 12.16 & 5.88 & 5.45 & 11.33 \\

  \bottomrule

\end{tabular}
}
\vspace{-18pt}
\end{table}

\vspace{-5pt}
\section{Results and Discussions}
\vspace{-5pt}

\subsection{Ablation Study}
\vspace{-3pt}

Table \ref{ablationtable} shows the ablation study results of our proposed FLCMA module and Pretrain strategy. To further represent the models' performance, we also evaluate the models by calculating the area under the receiver operating characteristic curve (AUC). The AV-Transformer/Conformer(E) uses the standard transformer/conformer encoder block without the FLCMA module. These models consist of a simple concatenation fusion module before the encoder blocks, which is described as $X_{fused}$ in Section \ref{sec:sec2}. The AV-Transformer/Conformer(L) uses the two split uni-modal transformer/conformer encoders without the FLCMA module, with a simple concatenation fusion module after the encoder blocks. AV-Transformer(E) and AV-Conformer(E) achieve a performance of 99.083\% and 99.231\% AUC, 8.95\% and 7.77\% WWS on the eval set, which shows that training from scratch using E2E training strategy can leverage the audio-visual multi-modal features at the same time and achieves good performance. AV-Transformer(L) and AV-Conformer(L) achieve a performance of 99.421\% and 99.416\% AUC, 6.67\% and 6.70\% WWS on the eval set. This suggests that fusing deep multi-modal latent features from the encoders is superior to fusing shallow features from the basic frontend module.

The FLCMA module enhances the performance of the E2E AV-Transformer/Conformer system. The FLCMA-based AV-Transformer achieves 99.518\% AUC and 7.06\% WWS on the eval set, outperforming the baseline AV-Transformer(E). Similarly, the FLCMA-based AV-Conformer achieves better 99.541\% AUC and 5.50\% WWS on the eval set. The result shows that the FLCMA module can model the correlation between modalities at the frame level through the high synchronous lip movements and speech signal. FLCMA based AV-Transformer is slightly worse than AV-Transformer(L) at the WWS score but is better at the AUC on the eval set, we find that the selection of thresholds easily influences the scores of WWS, and the classification ability reflected by AUC can be used as another reference. Compared to AV-Transformer/Conformer(L), FLCMA based AV-Transformer/Conformer achieves better AUC performance (+0.097/+0.125) on the eval set and shows the effectiveness of the FLCMA module when using a single parameter-less audio-visual multi-modal encoder instead of two split uni-modal encoders.


Finally, combined with the Pretrain strategy, both FLCMA based AV-Transformer/Conformer models improve their performance. Our final system (FLCMA-based AV-Conformer with Pretrain strategy) has a further 17\% reduction on WWS score, eventually reaching the WWS of 4.57\%. 

\begin{table}[t]
\small
\setlength{\tabcolsep}{1.35mm}{
  \centering
  \footnotesize
  \caption{\label{avtable} {\it Comparisons between pervious state-of-the-art audio-visual systems and ours. $*$ means the model with pretrain strategy.
  \vspace{+0.2em}
}}
  \begin{tabular}{ccccccccc}
  \toprule
  \multirow{2}*{\textbf{Model}}  & \multicolumn{3}{c}{\textbf{Dev[\%]}} & \multicolumn{3}{c}{\textbf{Eval[\%]}}  \\
  \cmidrule(lr){2-4} \cmidrule(lr){5-7}  & FRR & FAR & WWS & FRR & FAR & WWS \\
  
  \midrule
  
  Official \cite{zhou22g_interspeech}  & 7.3 & 6.8 & 14.1 & 10.1 & 15 & 25.1 \\
  Xu et al. \cite{beike} & - & - & - & - & - & 9.1 \\
  Cheng et al. \cite{cheng2022dku}  & 3.85 & 3.42 & 7.27 & - & - & 7.1 \\
  MISP 2021 1st \cite{mispchallenge}  & - & - & 4.1 & - & - & 5.8 \\
  \midrule
  Zhang et al. \cite{vekws}  & 1.60 & 2.02 & \textbf{3.62} & 2.79 & 2.95 & 5.74 \\
  Wang et al. \cite{dkuv2}  & 3.04 & 2.55 & 5.59 & 2.15 & 3.44 & 5.59 \\
    \midrule
  FLCMA-Transformer & 1.76 & 4.18 & 5.94 & 1.04 & 6.02 & 7.06 \\
  FLCMA-Transformer$^{*}$ & 2.24 & 2.21 & 4.45 & 2.14 & 3.33 & 5.47 \\
  FLCMA-Conformer & 2.24 & 2.31 & 4.55 & 1.59 & 3.91 & 5.50 \\
  FLCMA-Conformer$^{*}$ & 2.08 & 1.78 & 3.86 & 2.02 & 2.55 & 
\textbf{4.57} \\

  \bottomrule

\end{tabular}
}
\vspace{-18pt}
\end{table}

\vspace{-7pt}
\subsection{Results of the Pretrained single-modality model}
\vspace{-3pt}

Table \ref{audiotable} shows the performance of recent uni-modal systems on the MISP dataset. Our system outperforms the official baseline \cite{zhou22g_interspeech} and slightly underperforms the recent unimodal systems. We find that even though the uni-modal system for the multi-modal pretrained system does not achieve the best, it can further improve the performance of our multi-modal E2E system.

\begin{figure}[ht]
    \centering
    \vspace{-5pt}
    \includegraphics[width=0.38\textwidth]{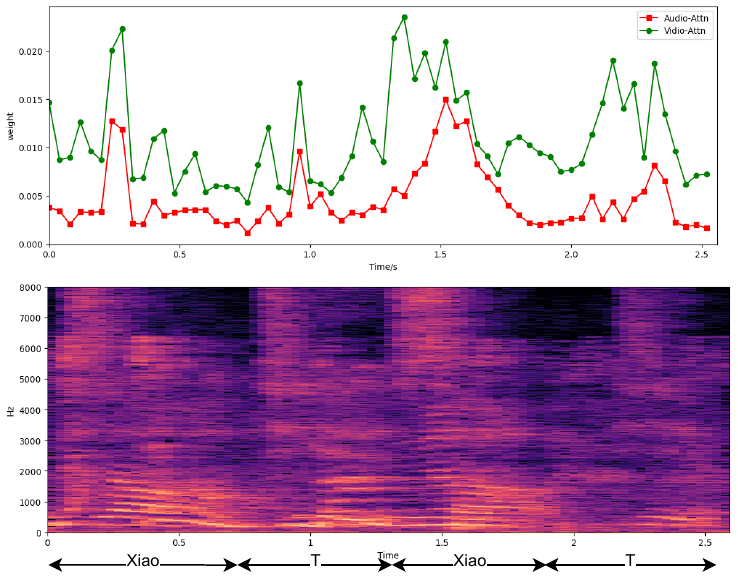}
    \vspace{-15pt}
    \caption{Visualization of the attention weights of the FLCMA based Audio-Viusal Conformer model.}
    \label{rolloutout}
    \vspace{-8pt}
\end{figure}

\begin{figure}[ht]
    \centering
    \includegraphics[width=0.38\textwidth]{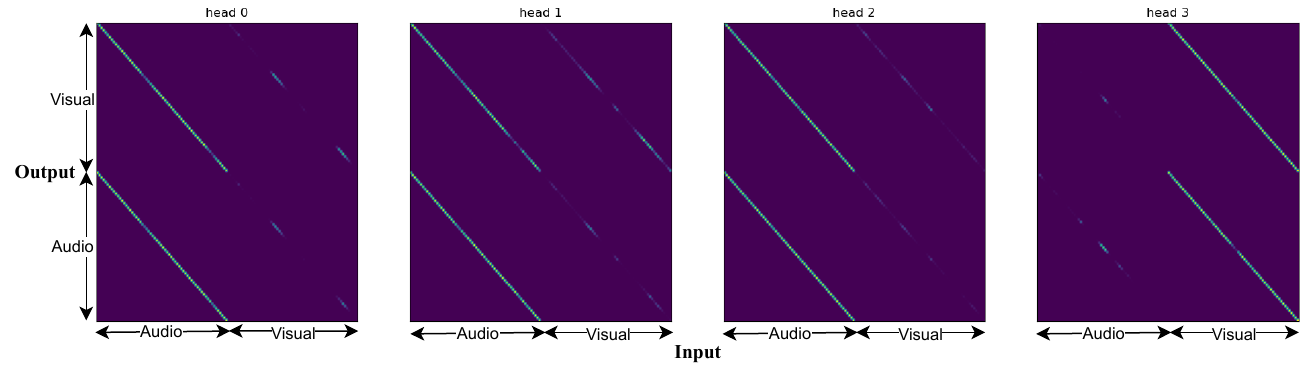}
    \vspace{-15pt}
    \caption{Visualization of attention maps in the FLCMA module of the last encoder conformer block. The head means the four head in the FLCMA module.}
    \label{lastattn}
    \vspace{-18pt}
\end{figure}

\vspace{-5pt}
\subsection{Results compared with previous works}
\vspace{-3pt}

Table \ref{avtable} compares the performance of our system with previous multi-modal approaches. 
\cite{dkuv2, vekws} first
train two robust single-modality models and then fuse these two frozen models together. Our FLCMA based Transformer with pretrain strategy and our FLCMA based Conformer achieve 5.47\% and 5.50\% WWS on the eval set, which are better compared with the previous works. Our model slightly underperforms \cite{vekws} on the dev set but outperforms it on the eval set. It does not overfit on the dev set, demonstrating better generalization performance. Finally, our FLCMA based Conformer with pretrain strategy achieves the SOTA audio-visual result (4.57\% WWS).

\vspace{-7pt}
\subsection{Visualization of the FLCMA module}
\vspace{-3pt}

We use the attention rollout \cite{abnar-zuidema-2020-quantifying} to visualize the attention weights of the FLCMA based Audio-Viusal Conformer model. Fig. \ref{rolloutout} illustrates higher attention weights around 0.25s, 1.0s, 1.5s, and 2.3s, corresponding to identifiable features of the wake word. We also visualize the attention maps in the FLCMA module of the last encoder conformer block. As shown in Fig. \ref{lastattn}, the features in the same modality have different attention weights at different frames. The model can customize attention weights at different frames, enabling it to capture inter-modality correlations. If one modality lacks confidence, the model can fuse information from multiple modalities for enhanced performance. 


%

\vspace{-10pt}
\section{Conclusion}
\vspace{-5pt}

In this work, we introduce the Frame-Level Cross-Modal Attention (FLCMA) module to enhance the performance of the Audio-Visual Wake Word Spotting system. This module enables the modeling of multi-modal semantic information at the frame level by leveraging synchronous lip movements and speech signals. We train the end-to-end FLCMA based Audio-Visual Conformer and further improve the performance by fine-tuning pre-trained uni-modal models. The proposed system achieves a new state-of-the-art result (4.57\% WWS score) on the MISP dataset, demonstrating the effectiveness of the approach.

\vspace{-10pt}
\section{ACKNOWLEDGMENTS}
\label{sec:copyright}
\vspace{-5pt}

This research is funded in part by the National Natural Science Foundation of China (62171207, 52105128), Science and Technology Program of Suzhou City (SYC2022051) and Youth Foundation Project of Zhejiang Lab (K2023BA0AA03). Many thanks for the computational resource provided by the Advanced Computing East China Sub-Center.




\bibliographystyle{IEEEbib}
\bibliography{strings,refs}

\end{document}